\documentstyle[12pt,epsfig]{article}
\def\lab#1      {\hbox{\small #1} }

\newcommand{\be}{\begin{eqnarray}}
\newcommand{\ee}{\end{eqnarray}}
\newcommand{\ben}{\begin{eqnarray*}}
\newcommand{\een}{\end{eqnarray*}}
\newcommand{\la}{\langle}
\newcommand{\ra}{\rangle}
\newcommand{\half}{\frac{1}{2}}
\newcommand{\pe}{\rightarrow}

\newcommand{\stb}{\alpha_{\hat{\partial} b}}
\newcommand{\stbi}{\alpha_{\hat{\partial} b_{i}}}
\newcommand{\stbp}{\alpha_{\hat{\partial} b'}}
\newcommand{\purr}{'}

\def\mb#1         {\mbox{\boldmath $#1$}}
\def\diffn#1	  {\Delta^{-}_{#1}}

\begin{document}

\begin{titlepage}

\begin{tabbing}
\` {\sl hep-lat/0002031} \\
    \\
\` LSUHE No. 341-2000 \\
\` February, 2000 \\
\end{tabbing}
 
\vspace*{0.1in}
 
\begin{center}
{\large\bf 
Vortices 
in $SO(3) \times Z(2)$ simulations
\\}
\vspace*{.5in}
A. Alexandru$^1$
and Richard W. Haymaker$^2$\\
\vspace*{.2in}
{\small\sl Department of Physics and Astronomy, Louisiana State
University,\\ Baton Rouge, Louisiana 70803-4001, U.S.A.
\\
\vspace{3ex}
e--mail addresses:      $^1$alexan@rouge.phys.lsu.edu, 
                  $^2$haymaker@rouge.phys.lsu.edu}
\end{center}

\vspace{0.2in}
\begin{center}
\begin{minipage}{5in}
\begin{center}
{\bf Abstract}
\end{center}

We study the configuration space of the Tomboulis $SO(3) \times Z(2)$ formulation with periodic boundary conditions. 
The dynamical variables are constrained by the required coincidence of 
Z(2) and SO(3) monopoles. We
propose an update algorithm that satisfies the constraints 
and is straightforward to implement.  We further prove
that this it reaches all configurations. We show how
the boundary conditions put constraints on the configuration space.
 We measure gauge invariant vortex counters for ``thin",
 ``thick" and ``hybrid" vortex sheets. For comparison we also 
measure projection vortex counters defined in the 
maximal center gauge.
\end{minipage}
\end{center}

\vfill
\noindent
PACS indices: 
11.15.Ha, 
11.30.Ly. 

\end{titlepage}

\setcounter{page}{1}
\newpage
\pagestyle{plain}

\section{Introduction}
\label{Introduction}

Lattice QCD continues to maintain an important role in the search
for the physics of color confinement.  The lattice regulator maintains gauge invariance
at all costs.  Dynamical variables are group elements rather than elements of a Lie algebra.
As a consequence many of the topological features that are prominent candidates 
for elucidating the physics of confinement have natural lattice definitions.  These include
U(1), Z(N), SU(N)/Z(N) monopole loops, Dirac sheets, Z(N) and SU(N) vortex sheets etc.  
These objects are often abundant in  U(1) and SU(N) lattice gauge theories.
They become singular only as one approaches the continuum limit.  

Consider the case of SU(N).  Further
consider a multiply connected region in which all links are gauge 
equivalent to $1$ on any simply connected patch of the region, i.e. all 
neighboring plaquettes   $ = 1$. 
Then the value of 
a Wilson loop lying in this region would take the value of a center element of
SU(N) $ = e^{2 \pi i n / N}$, which for $ n\ne 0$ indicates the trapping of a vortex. 
The occurance and absence of vortices gives a fluctuating
value that can disorder the Wilson loop and lead to an  area law.  

Yaffe\cite{y}, Tomboulis\cite{t} and Kovacs and Tomboulis\cite{kt}
have developed a formulation of SU(N) gauge theory that is manifestly   SU(N)/Z(N) 
invariant.  In this formulation, center elements, Z(N), multiplying each link leave the
action and measure invariant. New Z(N) variables, defined on 
plaquettes, $\sigma(p)$ ,  carry the Z(N) degrees of freedom.
This  formalism, equivalent to the standard SU(N) form,  
allows an elegant topological 
classification of the SU(N)/Z(N) and Z(N) vortex 
configurations occuring on the lattice.  See also 
references [4-7].

In this paper, we address the issues of simulating the Tomboulis formulation on a periodic
lattice.  Many results have followed from this formulation without doing simulations in these
variables\cite{ttt}.  As a first calculation, we tag Wilson loop measurements by the occurance 
of vortices linking the loop using a number of different vortex counters.  We 
restrict our attention to SU(2) in this paper.

We also measure the P (projection) vortex counter in the original SU(2) formulation 
following [9-17] 
for comparison.   Projection vortices arise in a Z(2) gauge
theory derived from the original SU(2) theory by going to the maximal center gauge  and then replacing links  by 
$\lab{sgn} ( \lab{tr} ( U_{x, \mu}))$.
The projected theory has ``thin" Z(2) vortices defined on one lattice
spacing.  They have been found to be well
correlated with center vortices and therefore a 
measurement of P vortices is a predictor of them.

The thrust of this paper is to study the configuration space in the Tomboulis
variables on the torus.  These variables are subject to constraints in a rather indirect way.    
We propose a constructive update algorithm which is straightforward to implement and
we prove that it reaches all configurations, i.e. it is ergodic.

In section 2 we rederive the Tomboulis
form of the partition function. There are configurations on the torus that give zero
weight which we exhibit in section 3.  In section 4 we
elucidate two alternative definitions of the configuration space for the Z(2) variables
 $\sigma(p)$, the indirect definition and constructive definition.  
Appendix B gives the proof of their equivalence.  
In section 4, we discuss various vortex counters.
In section 5, we measure 
these vortex counters for Wilson loops.  This allows us to 
tag Wilson loops to study the disordering mechanism.
In Appendix C we consider the case of anti-periodic boundary conditions\cite{private}.

\section{Derivation of $Z_{Z(2) \times SU(2)/Z(2)}$ on a torus }
\label{Z}

\underline{}
Our derivation here is due to Tomboulis \cite{t}, and reviewed recently by Kovacs and 
Tomboulis \cite{kt} for free boundary conditions. The latter paper gives a 
thorough pedagogical review of the formulation and the topological
features.   In addition they have paid close attention to 
visualization of vortices, both as surfaces and 
as curves in 3-d slices.  

Consider the partition function for the Wilson action,
\be
Z 
=
\int 
\left[
  d U(b) 
\right]
\exp
\left(
   \beta 
   \sum_{p}
   \half
   \lab{tr} [U(\partial p)]
\right).
\label{wilsonaction}
\ee
Define a Z(2) variable on the links (bonds), $\gamma(b)$, and insert
a constant into the partition function,
\be
Z  
= 
\sum_{\gamma(b)}
\int 
\left[
   d U(b) \gamma(b)
\right]
\exp
\left(
   \beta 
   \sum_{p} 
   \half \lab{tr} [  U(\partial p)]  
\right).
\label{z2links}
\ee
(See Appendix A for Z(2) notation, algebra, characters, delta functions, etc.)
Apply the Haar invariant transformation
\ben
U(b) \rightarrow U(b) \gamma(b); \;\;\;\;\;\;\;\;
Z  
= 
\sum_{\gamma(b)}
\int
\left[
     d U(b) 
\right]
\exp
\left(
   \beta 
   \sum_{p} 
   \half \lab{tr} [  U(\partial p)] 
   \gamma(\partial p)
\right).
\een
Isolate the sign of the plaquette, $\eta(p)$, 
$\lab{tr} [U(\partial p)] = |\lab{tr} [U(\partial p)]| \times \eta(p)$, writing
\ben
Z  
= 
\sum_{\gamma(b)}
\int
\left[
     d U(b) 
\right]
\exp
\left(
   \beta 
   \sum_{p}
   \half|\lab{tr} [  U(\partial p)]| 
   \gamma(\partial p) 
   \eta(p)
\right).
\een
Next introduce a new Z(2) variable, $\sigma(p)$, defined on plaquettes, 
\ben
1  
= 
\sum_{\sigma(p)}
   \delta( \sigma(p) \times \gamma(\partial p) \eta(p)),
\een
\ben
Z =
\sum_{\gamma(b)}
\int 
\left[
     d U(b) 
\right]
\left\{
\sum_{\sigma(p)}  \delta( \sigma(p) \times \gamma(\partial p) \eta(p))
\right\}
\exp
\left(
   \beta \sum_{p} \half|\lab{tr} [  U(\partial p)]| \sigma(p)
\right).
\label{z21}
\een
Expand the delta function in Z(2) characters
\ben
\delta( \sigma(p) \eta(p)\times \gamma(\partial p))
&=&
\half \sum_{\tau(p)} 
\left\{
   \chi_{\tau(p)}\left(\sigma(p) \eta(p)\right) \; 
   \chi_{\tau(p)}\left(\gamma(\partial p)\right)
\right\},
\nonumber
\\
&=&
\half \sum_{\tau(p)} 
\left\{
   \chi_{\tau(p)}\left(\sigma(p) \eta(p)\right) \; 
   \chi_{\gamma(\partial p)}\left(\tau(p)\right)
\right\}.
\een
This gives
\ben
Z  
&=&
\sum_{\gamma(b)}
\int 
\left[
  d U(b)
\right]
\sum_{\sigma(p)}
\sum_{\tau(p)}
\nonumber 
\\ 
&& 
\left\{
   \prod_{p}  \half
   \chi_{\tau(p)}\left(\sigma(p) \eta(p)\right)
\right\}
\left\{
   \prod_{p}
   \chi_{\gamma(\partial p)}\left(\tau(p)\right)
\right\}
\exp
\left(
   \beta 
   \sum_{p} 
   \half|\lab{tr} [  U(\partial p)]| 
   \sigma(p)
\right). 
\label{z33}
\een
In the second character, we can rearrange the product over plaquettes    
into a product over links,
\ben
   \prod_{p}
   \chi_{\gamma(\partial p)}\left(\tau(p)\right)
 = 
   \prod_{b}        
   \chi_{\gamma(b)}\left( \tau(\widehat{\partial} b ) \right).
\label{z35}
\een
Now do the $\gamma(b)$ summation
\be
Z  
&=&
\int 
\left[
     d U(b)
\right]
\sum_{\sigma(p)}
\nonumber 
\\ 
&& 
\left[
   \sum_{\tau(p)}
   \prod_{p}  
   \half
   \chi_{\tau(p)}\left(\sigma(p) \eta(p)\right)
   \prod_{b}        
   \delta \left( \tau(\widehat{\partial} b) \right)
\right]
\exp
\left(
   \beta 
   \sum_{p} 
   \half|\lab{tr} [  U(\partial p)]| 
   \sigma(p)
\right) .
\label{z36}
\ee

The item in square brackets is the starting point for much of the analysis in this
paper. The dynamical variables  are 
$\{ U(b), \sigma(p)\}$. The $\tau(p)$ sum can be done.  
\be
C(\sigma(p)\eta(p)) 
&\equiv& 
\left[
   \sum_{\tau(p)}
   \prod_{p}  
   \chi_{\tau(p)}\left(\sigma(p) \eta(p)\right)
   \prod_{b}        
   \delta \left( \tau(\widehat{\partial} b) \right)
\right],
\label{tau}
\\
&=&
\prod_{c}
\delta\left(\sigma(\partial c) \eta(\partial c) \right)
\times
\left\{
\begin{array}{c}
1\\
0
\end{array}.
\right\} \times \lab{constant}
\label{C}
\ee
The constraint $\delta \left( \tau(\widehat{\partial} b) \right)$ means that there must
be an even number of $\tau = -1$ plaquettes in the co-boundary of the links 
(i.e. the six plaquettes contiguous with the link).

The last equality needs further explanation. If a cube contains an odd number of
faces with $\eta = -1$ then by definition 
it contains an SO(3)  monopole. Similarly if a cube
contains an odd number of faces with $\sigma = -1$ then 
it contains a Z(2) monopole.  
The delta function on the cube requires
that any SO(3) monopole be paired with a Z(2) monopole at the same location.
We show in Sec. 4 that for the integrand of the partition function 
 to be different from zero it is necessary to have 
$\prod_{c}  \delta\left(\sigma(\partial c) \eta(\partial c)\right)=1$.
However on the torus, this is not sufficient.  
There are configurations, 
$\left\{U(b), \sigma(p) \right\}$,
for which the delta functions on the cube
are unity  yet the integral  vanishes.  We denote these configurations
`` weight = 0."
In section \ref{zero_weight}
we give examples of such  configurations
which are co-closed vortex 
sheets that wrap around periodic boundary conditions.
We further show that when the $\tau$ sum differs from zero, it is 
a constant in the  variable 
\be
\alpha(p) \equiv  \sigma(p) \eta(p).
\label{alpha}
\ee
We denote these  ``weight $= 1$" configurations.
In section \ref{simulation} we exhibit 
an update algorithm that reaches
all weight $= 1$ configurations and respects all constraints.

Restricting $\left\{U(b) \right\}$
and $\left\{\sigma(p) \right\}$, to the weight $= 1$ configurations
we obtain
\be
Z  
= 
\sum^{'}_{\sigma(p)}
\int 
\left[
    d U(b)
\right]'
\prod_{c}
 \delta\left(\sigma(\partial c) \eta(\partial c)\right)
\exp
\left(
   \beta 
   \sum_{p} 
   \half|\lab{tr} [  U(\partial p)]| 
   \sigma(p)
\right) .
\label{z37}
\ee
Note that this form is invariant under
$U(b) \rightarrow \gamma(b) U(b)$.   All configurations related by this 
transformation are SU(2) representatives of the invariance group SO(3). 
The $Z(2)$ part is explicit in the $\sigma(p)$ variables.

\section{Zero weight configurations on the torus}
\label{zero_weight}

In this section we find configurations on the torus which have zero
weight as indicated in Eqn.(\ref{C}). We take periodic boundary 
conditions in all directions. The zero weight reflects the fact that a 
vortex wrapped around the torus is topologically stable\cite{private}.  It can not be reached
from a non-vortex configuration.  This property is inherent in the
formalism.

Our starting point is Eqn.(\ref{tau}). Using Eqn.(\ref{alpha})
this becomes
\be
C(\alpha(p)) 
= 
\sum_{\tau(p)}
   \prod_{p}  
   \chi_{\tau(p)}\left(\alpha(p) \right)
   \prod_{b}        
   \delta \left( \tau(\widehat{\partial} b \right).
\label{CC}
\ee
The $\tau$ delta function constraint requires that the plaquettes forming 
the co-boundary of any link must occur in even numbers.  Clearly
a closed surface made by tiling with $\tau = -1$ plaquettes, with  
$\tau = +1$ elsewhere, will satisfy all these constraints.

Now lets turn to the $\alpha$ variables.
Consider a configuration in which $\alpha(p)= -1$ on all plaquettes
$p^{1 2}_{0 0 k \ell}$ and  $ = +1$ elsewhere.  
The upper indices denote the plaquette orientation, 
and the lower indices are the space-time coordinates, $(i, j, k, \ell)$. 
This configuration is a co-closed vortex sheet wrapped 
around the torus in the $3$ and $4$ directions.  
If $\sigma(p) = \mp 1$ and $\eta(p) = \pm 1$ everywhere on this co-closed vortex, 
it is a $\sigma$/$\eta$ vortex. 
If these two cases occur on different patches of the vortex sheet, 
then it is a hybrid vortex with co-closed monopole loops at the
boundaries between the patches. 

The $\alpha$ cube delta function constraints, Eqn.(\ref{C}), are 
satisfied because cubes will either have
no vortex plaquettes in common, 
or will have two vortex plaquettes on opposite faces.
In spite of this we now show that this $\alpha$ configuration 
has zero weight in the partition function.   

Consider Eqn.(\ref{CC}) applied to an arbitrary function of 
$F(\tau(p),\alpha(p))$. Define the set $\cal C$ 
\ben
\sum_{\tau(p)}  F(\tau; \alpha)
   \prod_{b}        
   \delta \left( \tau(\widehat{\partial} b \right)
 \equiv \sum_{\cal C} F(\tau; \alpha),
\een
i.e. it is the set of all $\tau$ configurations with the property that 
$  \prod_{b}        
  \delta \left( \tau(\widehat{\partial} b) \right)
=1$.

The set ${\cal C}$ forms a group under the multiplication defined 
through
\be
(\tau_1 \tau_2)(p) = \tau_1(p) \tau_2(p).
\label{group}
\ee
Given that $\tau_1$ and $\tau_2$ are group elements then each has an even number of negative plaquettes in the co-boundary of any link. 
Clearly the product will have the same property. The identity element is 
$\tau(p)= 1$  for all $p$, and the elements are their own inverses.

Using the invariance property of the group summation:
\ben
\sum_{\tau\in\cal C} F(\tau_0\tau; \alpha) = 
\sum_{\tau\in\cal C} F(\tau; \alpha),
\een
where $\tau_0$ is any element of the group $\cal C$.

Substituting for $F$ our case reads:
\be
C(\alpha(p)) 
&=& 
\sum_{\tau(p)}
   \prod_{p}  
   \chi_{\tau(p)}\left(\alpha(p) \right)
   \prod_{b}        
   \delta \left( \tau(\widehat{\partial} b )\right), 
\label{CCC}
\\
&=& 
\sum_{\tau(p)}
   \prod_{p}  
   \chi_{\tau_0(p) \tau(p)}\left(\alpha(p) \right)
   \prod_{b}        
   \delta \left( \tau(\widehat{\partial} b )\right), 
\nonumber
\\
&=&
   \prod_{p}   
   \chi_{\tau_0(p)}\left(\alpha(p)\right)
 \times C(\alpha(p)),
\label{zeroo}
\ee
for any $\tau_0\in\cal C$. Therefore if we can find a group element
$\tau_0$ for which $ \prod_{p}      \chi_{\tau_0(p)}\left(\alpha(p)\right)
 \ne 1$ then $C(\alpha)  = 0$.

Choose 
$\tau_0(p) = -1$ on the subset of plaquettes: $\{p_{i j 0 0}^{12}\}$ for all $i, j$ and +1 elsewhere. It is obviously a member of 
the set ${\cal C}$. This is a closed tiled surface 
of $1,2$ plaquettes wrapping around the $1,2$ directions
for the $3$ and $4$ coordinates fixed  to  $0$.
Using the
fact that $\chi_+(+) = \chi_+(-) = \chi_-(+) = +1 = - \chi_-(-)$,
we count the number of sites where both $\alpha = -1$ and 
$\tau_0 = -1$.   

The $ \alpha(p) $ configuration under consideration is a co-closed set of $1,2$ plaquettes
wrapped around the $3,4$ directions and the $ \tau(p) $ configuration  is a closed
tiled set of $1,2$ plaquettes wrapped around the $1,2$ directions.  They have only one negative
plaquette in common at position $(0,0,0,0)$.  
Therefore $ \prod_{p}      \chi_{\tau_0(p)}\left(\alpha(p)\right)
 = -1$ implying $C(\alpha) = 0$.

\section{$\sigma(p)$ configuration space}
\label{simulation}

We have seen examples in the last section of the interplay between the $\tau$ and
$\alpha$ configurations in finding non-zero contributions 
to the partition function.  We are interested in 
simulating in the variables $\{U(b), \sigma(p) \}$.  
Thus far the allowable $\alpha \;( = \sigma \times \eta)$
configurations are defined indirectly in terms in the allowable $\tau$ configurations
which are also indirectly defined by constraints.   The corresponding simulation would
also be very indirect and perhaps difficult to implement.

We propose a constructive definition of allowable $\alpha(p)$ configurations by building
them up from ``star transformations", i.e. correlated sign flips of the $\sigma $ plaquettes 
occurring in the co-boundary of each link. 
Since these are constrained updates of six plaquettes, it is not clear that we can reach
all allowed  $\{ \sigma \}$ 
(or equivalently  $\{ \alpha \}$) configurations  by this method.  
However we show that this definition of allowed
$\alpha(p)$ configuration is identical to the above definition.  The proof is relegated
to Appendix B.  In this section we give a summary of the result.

Before discussing the $\sigma$ configurations let us first describe the link updates.
This is a straightforward generalization of the link updates for $SU(2)$.   
The proposed
change in a link might change the sign of the $\eta$ plaquettes in the co-boundary of 
the link.  If one of these changes sign, we need to flip the sign of the 
corresponding $\sigma$ plaquette so that the $\alpha$ configuration is unchanged.
Then the Monte Carlo step is essentially the same as for the SU(2) update. 

Next consider the above mentioned ``star transformations".   Our proposed 
update is to flip the sign of the six $\sigma$ plaquettes forming the co-boundary of 
the links.

Assume we are starting from a weight $= 1$ configuration, Eqn.(\ref{C}). It is easy 
to see that both these update steps will preserve the cube constraints. 

In the previous section we described vortex sheets that wrap around the torus. Consider 
the operator constructed out of $\mu, \nu$ plaquettes\cite{kt}
\ben
{\cal N}_{\mu, \nu}  =  \prod_{p\in S_{\mu, \nu}} 
\eta(p) \sigma(p)  \equiv  
 \eta(S_{\mu, \nu})  \sigma(S_{\mu, \nu}) = \pm 1,
\een
where $S_{\mu, \nu}$ is a whole tiled $\mu, \nu$ plane.  
${\cal N}_{\mu, \nu} = \pm$ 
for an even/odd number of vortices of stacked $ \mu, \nu $ plaquettes wrapping 
around the orthogonal $\xi, \eta$ directions of the torus. 

We start the algorithm with this 
${\cal N}_{\mu, \nu} = +1$  in all 6 planes. 
It is easy to see that our update algorithm preserves 
${\cal N}_{\mu, \nu}$.
Hence, starting with non-zero weight configurations, we do not generate the zero
weight configurations described in the last section, since this would involve
$\Delta {\cal N}_{\mu, \nu} \ne 0$.

Let us define the relevant sets of configurations more carefully.
Consider Eqn.(\ref{CCC})
\be
C(\alpha) 
&\equiv& 
\sum_{\tau(p)}
   \prod_{p}  
   \chi_{\tau(p)}\left(\alpha(p)\right)
   \prod_{b}        
   \delta \left( \tau(\widehat{\partial} b) \right), 
\nonumber
\\
&=& 
\sum_{\tau \in {\cal C}}
   \prod_{p}   
   \chi_{\tau(p)}\left(\alpha(p)\right),
\nonumber
\\
&\equiv& 
\sum_{\tau \in {\cal C}}
\la  \tau, \alpha \ra \;\;\; = \;\;\;
\sum_{\tau \in {\cal C}}
\la  \alpha, \tau \ra.
\label{CCCC}
\ee

The third line is a shorthand for the
product over characters (see Appendix B). 
The bracket, $\la  \alpha, \tau \ra = \pm 1$, is negative if and only if 
there are an odd number of plaquettes for which 
both $\alpha(p) = -1$ and $\tau(p) = -1$.

The second line is an alternative way to specify the sum, where:
\ben
{\cal C}
= 
\{\tau\in{\cal A}| \prod_{b} \delta \left( \tau(\widehat{\partial} b) \right)=1\}.
\een

Configurations form a group, Eqn.(\ref{group}).  
${\cal C}$ is a subgroup of the group  ${\cal A}$
of all configurations,  $|{\cal A}| = 2^{6 N}$ in number, where
$N$ is the number of lattice sites. Restating:
\begin{quote}
{\em ${\cal C}$ is the group of all configurations $\{\tau\}$ with an
 even number of $\tau = -1$ 
plaquettes 
occurring in the co-boundary of every link, i.e. 
forming a closed tiled  surface  of negative plaquettes.} 
\end{quote}

There is a second group of interest, 
\ben
\overline{\cal C}=\{\alpha \in{\cal A}|\la\alpha,\tau \ra 
=1 \;\;\;\;\forall\tau\in{\cal C}\}.
\een
Restating:
\begin{quote}
{\em
${\overline{\cal C}}$ 
is the group of all configurations 
$\{ \alpha \}$ 
for which
$C(\alpha)$ may be different from zero.  We will show that on this set we have indeed $C(\alpha)\ne 0$. (See appendix B.2, Proposition 2)
Recall from  Eqn.(\ref{CC}) that if we can find a single configuration
$\tau_0$ 
for which 
$\la \alpha, \tau_0  \ra \ne 1$ 
 then 
$C(\alpha) = 0$.
}
\end{quote}

Therefore ${\overline{\cal C}}$ is the group of $\alpha$ 
configurations which have weight $=1$
in the sense of Eqn.(\ref{C}), i.e. the configurations that contribute to the partition
function.  Further,  ${\cal C}$ is the group of $\tau$ configurations that form 
closed tiled surfaces as required by the explicit constraints in  Eqn.(\ref{CCCC}).
In the previous section, we found a zero weight configuration $\alpha$ by finding
a configuration $\tau_0$ for which $\la \alpha, \tau_0 \ra = -1$.

The group ${\cal C}$ has only an implicit definition here.  The group ${\overline{\cal C}}$
has an implicit definition in terms of this ${\cal C}$. Therefore its definition 
is even more indirect.  Even without an explicit definition,
we have been able to specify precisely those configurations $\{ \alpha \}$ that contribute
non-zero weight to the partition function.   

There is a third group of interest,
\ben
{\cal D}=\{\alpha\in{\cal A}|\alpha=\prod \stbi \}.
\een 
where $\stbi$ refers to an individual star transformation on the 
$i$'th link $b_i$, and the product indicates
all possible products of them.

Restating:
\begin{quote}
{\em
${{\cal D}}$ 
is the group of all configurations 
$\{ \alpha \}$ 
which can be built out of products 
of ``star transformations" starting from the
identity configuration.
}
\end{quote}

This is the constructive definition that is straightforward to implement in a simulation.
The proof in Appendix B shows that the group $\cal D$ is identical to the group
${\overline{\cal C}}$.  In this way we have shown that by our proposed algorithm
is ergodic, reaching all configurations allowed in Eqn.(\ref{C}).

Let us return to Eqn.(\ref{C}) and the ``cube constraints."  Using our definition, $\alpha(p)=\sigma(p)\eta(p)$, the constraints are written as $\prod_{c} \delta(\alpha({\partial}c))=1$.  The cube constraint simply 
asserts that for a configuration to give a non-vanishing contribution,  
every cube in the lattice must have an even number of faces with $\alpha=-1$.

Let us suppose that a particular cube has an odd number of faces
 with $\alpha = -1$.  Then consider a configuration $\tau_0$ 
which takes values $-1$ on all 6 faces of this particular cube.  This is a closed
surface and therefore satisfies the constraints imposed on $\tau$.  For this case
\ben
   \prod_{p}   
   \chi_{\tau_0(p)}\left(\alpha(p)\right)
 = -1
\een
and therefore by Eqn.(\ref{zeroo}) $C(\alpha) = 0$.

The reason for signaling out this {\em necessary} constraint is that it is local. 
The zero weight configurations described in the last section necessarily wind around
the torus.

\section{Simulation of  vortex counters}

Vortices have long been considered as prime candidates for the
essential dynamical variable to describe confinement.    A simulation offers
a tool that allows one to correlate the occurance of vortices with values of other dynamical variables. Hence as 
a first application we use this formalism to measure various vortex counters for
Wilson loops.

Consider the SU(2) formalism with the standard Wilson action.  Further
consider a multiply connected region in which all links are gauge 
equivalent to $1$ on any simply connected patch of the region, i.e. all 
plaquettes  $ = 1$. This could occur if the vortices are very dilute and have a cross section  small compared to average separation.  

Then the value of 
a Wilson loop lying in this region  $= \pm 1$, the   center of SU(2),
 corresponding to an even/odd number 
of SU(2) vortices linking the region.  
 The occurance and absence of vortices gives a fluctuating
value that can disorder the Wilson loop and lead to an  area law.

In the $SO(3) \times Z(2)$ formulation the Wilson loop is given by 
\ben
W[C] =  \la \half \lab{tr} [C] \eta_S \sigma_S \ra_{C = \partial S},
\label{wilson}
\een
where $\eta_S$ and $\sigma_S$ are products of $\eta$ and $\sigma$ over any 
spanning surface\cite{y,t,kt}. 

Kovacs and Tomboulis\cite{kt} define three  vortex counters for 
thick vortex sheets, thin vortex sheets, and hybrid (patches 
of each on the sheet). Their definitions require measurements on {\em all} spanning surfaces.  
We measure here only the minimum spanning surface. 
Hence we must interpret our measurements as best we can in this limited simulation.
\begin{itemize}
\item Thin:
\ben
{\cal N}_{\lab{thin} } \equiv  \sigma_{S}.
\een
If this value, $ = \pm 1$, is independent of the spanning surface, then 
this counts thin vortices.
\item  Thick:
\ben
{\cal N}_{\lab{thick} } \equiv \eta_{S} \; \lab{sgn}[tr W(C)].
\een
This object is counting something more elusive since unlike the above case, the  vortex structure is spread over many lattice spacings.  
Nevertheless 
it is always possible to find a representative of SO(3) such
that the $\eta$  vortex defines the 
topological linkage\cite{kt, private}. 
The  $\eta$  
vortices can be deformed by a $Z(2)$ transformation of links 
giving  different representatives of SO(3) without cost
of action.  
One can move a linked $\eta$ vortex sheet so that it no 
longer links the Wilson loop and further even transform it away.  
However in this case the negative contribution will be transferred
to one of the perimeter links  of the Wilson loop, and it will not affect the value of  ${\cal N}_{\lab{thick} }$.  Again if this is independent
of the spanning surface, then this counts  ``thick" vortices.
\item  Hybrid:
\ben
{\cal N}_{\lab{hybrid} } = {\cal N}_{\lab{thin} } \times
{\cal N}_{\lab{thick} } =   \sigma_{S} \eta_{S} \;\lab{sgn}[tr W(C)].
\een
As one considers all spanning surfaces, the sign of 
${\cal N}_{\lab{thin} }$ might change. However
if the sign of ${\cal N}_{\lab{thick} }$ always compensates then 
this counts hybrid vortices.

\end{itemize}

We measure these three counters
 for Wilson loops, taking the
minimal spanning surface.  Then, for example, 
${\cal N}_{\lab{thin} } = -1$ does not distinguish
thin from hybrid, and similarly ${\cal N}_{\lab{thick} } = -1$ does not distinguish thick from 
hybrid. However they do measure the occurance of an object piercing the spanning surface
responsible for sign fluctuations of the Wilson loop.  Notice that ${\cal N}_{\lab{hybrid} }$
is just the sign of the Wilson loop, Eqn.(\ref{wilson}), and can be calculated in the
original SU(2) formulation.

A fourth definition of vortex counter enters in the work
of Del Debbio, Faber, Greensite and Olejnik\cite{dfgo}.  
See also Refs.(\cite{fgo,agg,lr,lrt,elrt,afe})    Starting with the SU(2) 
formalism, go to the maximal center gauge which maximizes
\ben
\sum_{x, \mu} |\lab{tr}( U_{x, \mu})|.
\een
Then consider a $Z(2)$ gauge theory in which the links are replaced by their $Z(2)$ values
\ben
\lab{sgn} ( \lab{tr} ( U_{x, \mu})).
\een
Denote the plaquettes in this Z(2) gauge theory  by $\xi(p)$.
The negative plaquettes of this theory, $\xi(p) = -1 $ form thin vortices, 
residing on one lattice spacing similar to $\sigma$ vortices.  They are denoted
P (projection) vortices. The authors find evidence for a strong correlation
between P vortices and thick objects, denoted center vortices, analogous to the thick
vortices of Tomboulis. Their calculations proceed in the original SU(2) theory with
the added observable  of the  P vortex counter to segregate contributions
to Wilson loops.  
\begin{itemize}
\item  Projection:
\ben
{\cal N}_{\lab{projection} } = \xi_{S}.
\een
This object is independent of the spanning surface.
\end{itemize}

We also measure the P vortex counter for comparison, using the SU(2) formalism.

\section{Numerical Results}

Simulations were done on a $12^4$ lattice for $\beta = 2.30$. 
 Measurements were binned
to 10 bins and jackknife errors calculated.
[thin]: 200 ;
[thick]: 400 ;
[hybrid]: 200 ;
[projection]: 1000 measurements.
We monitor the coincidence of Z(2) and SO(3) monopoles which can
slip due to round off error.

Let us emphasize again that in our application here 
these counters measure 
only the occurance of various objects 
piercing the minimal spanning surface, not the 
species of vortex. More general analysis will be given elsewhere\cite{ach}.

\begin{figure}
\epsfig{file=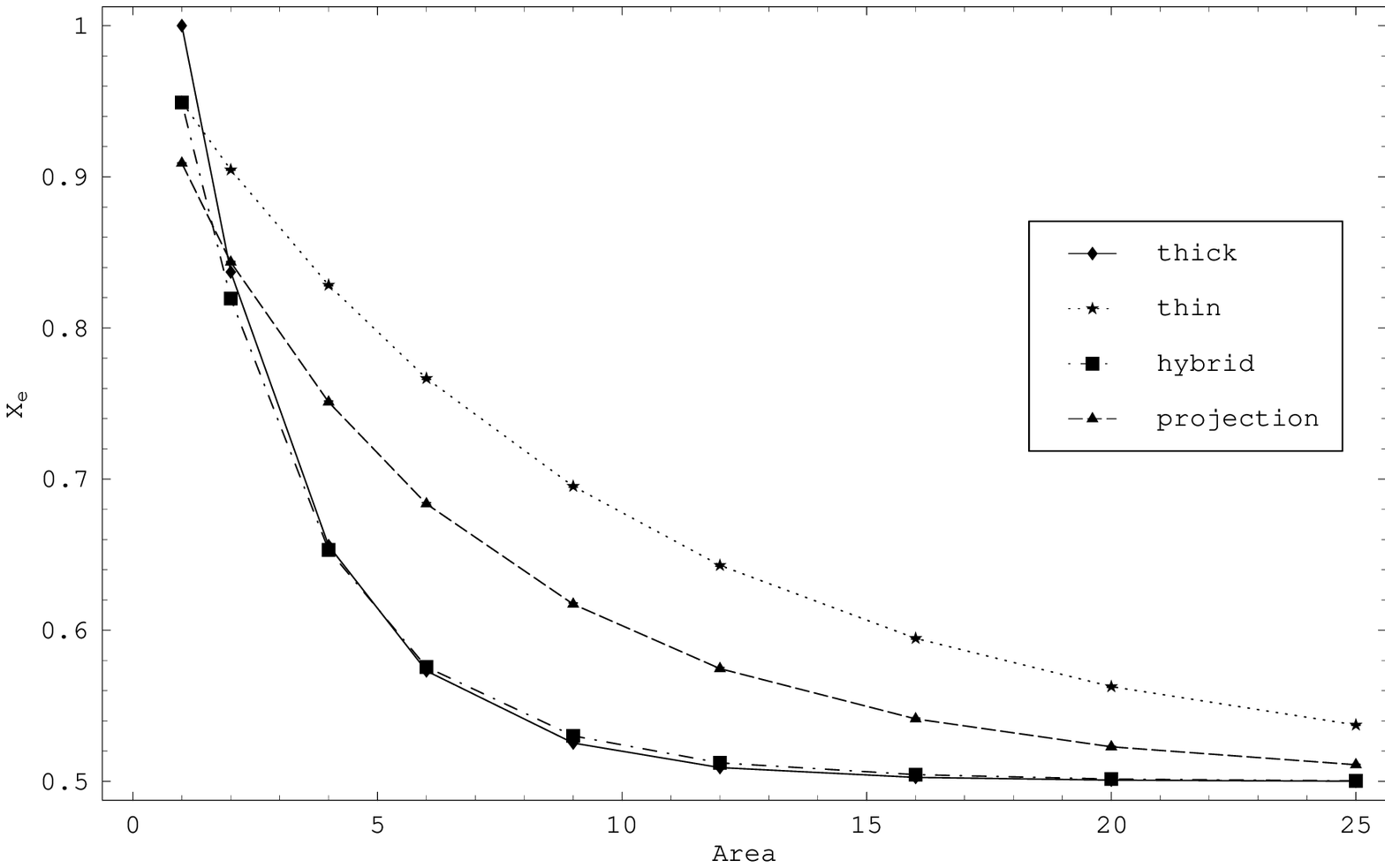,width=15cm}
\center{{\bf Figure 1}: Fraction of Wilson loops with an even number of
vortices piercing the minimal surface, $x_e$. ($x_o=1-x_e$)}
\end{figure}

Fig. 1 shows the fraction of Wilson loops 
which have an even number of vortices, $x_e$,
as a function of the area.
All the counters 
approach $50 \%$ from above with similar behavior ($x_o = 1 - x_e$).
They are each counting different things and 
are not expected to be equal.   

${\cal N}_{\lab{hybrid} }$ is a good reference curve since it 
measures the sign of the Wilson loop itself.  The area law arises from a  near cancelation of fluctuating values due to approximately 
equal occurance and absence of thin, thick or hybrid vortices.

The ``thick" curve lies on top of the ``hybrid" one. 
Hence the added factor of  $\sigma_S$ in the hybrid counter has little effect here.  We expect $\sigma$ vortices to be heavily suppressed
for increasing $\beta$ since they cost action proportional to the vortex area.   However at $\beta = 2.30$ the density of Z(2) (or SO(3)) monopoles 
$ = 0.2155(2)$  (Random plaquette signs would give a density of $ ~ 0.5$). 
Hence in spite of the near coincidence of these two curves, $\sigma$ vortices  and $\sigma$ patches of hybrid vortices  are important at this value of  $\beta$.  There is further evidence below.

Since Wilson loops are positive, one expects that the fraction of even loops $x_e$ should always dominate, as they do in all cases.

There are two special points on these curves. 
${\cal N}_{\lab{thick} } = +1$ by definition for a plaquette,
i.e. a  single ``thick" vortex can not
link a $1 \times 1$ Wilson loop.  Hence the fraction $x_e = 1$.  
Further this  gives the fraction $x_e$ equal 
for the two cases  ${\cal N}_{\lab{thin} }$, and 
${\cal N}_{\lab{hybrid} } $

The ${\cal N}_{\lab{projection} } $ 
case gives the same result as reported in Ref \cite{dfgo}.

Figs 2 - 6  shows Wilson loops, $W$, and the tagged Wilson loops corresponding
to even: $W_e$ and odd:  $W_o$ number of vortices as a function of 
loop area and their logarithmic derivatives.

\begin{figure}
\epsfig{file=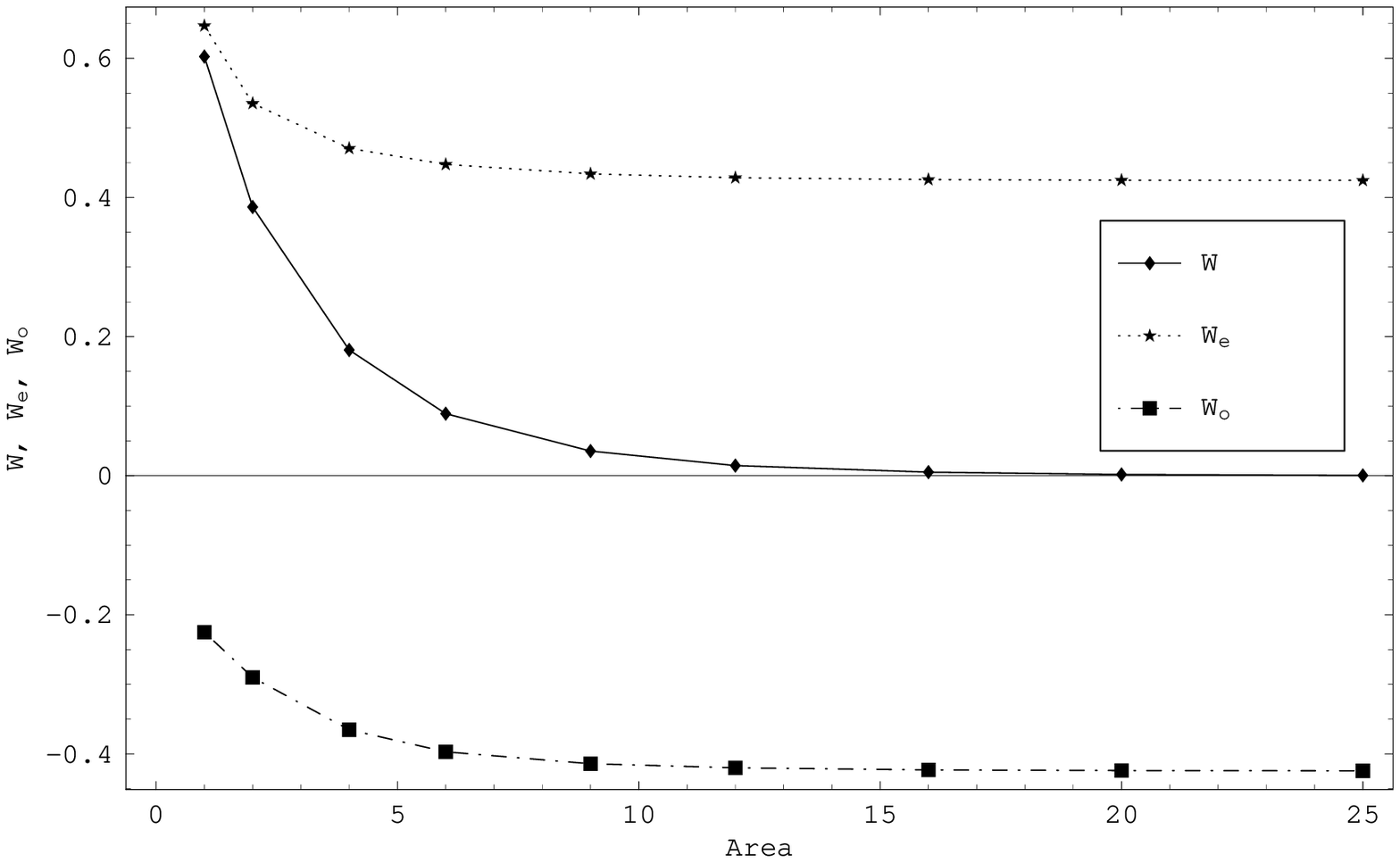,width=15cm}
\center{{\bf Figure 2}: Wilson loop, W, and tagged Wilson loop with even or 
odd number of ``thin" or ``thick" vortices piercing the minimal surface.}
\end{figure}

\begin{itemize}
\item Fig. 2: 
[Thick or thin segments piercing the minimal surface]
This vortex counter is just the sign of the Wilson loop itself.  
The positive and negative contributions to the Wilson
loop are averaged separately. Since each component has approximately
an equal and opposite asymptote, this illustrates the cancellations
due to disordering and the difficulty of measuring large loops.
\begin{figure}
\epsfig{file=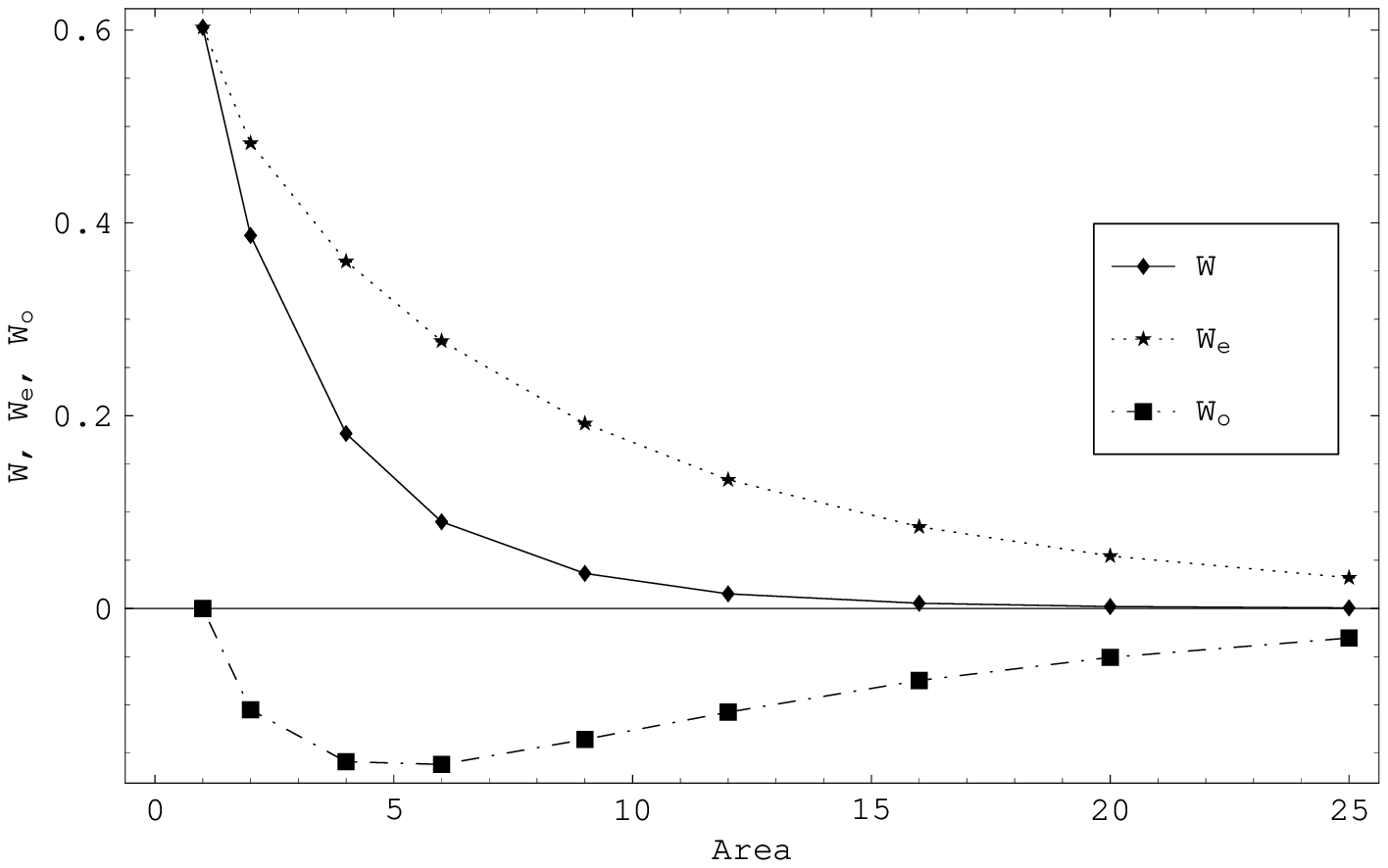,width=15cm}
\center{{\bf Figure 3}: Same as in Figure 2 but tagged by ``thick" only.}
\end{figure}

\item Fig 3: 
[Thick segment piercing the minimal surface] 
By definition $W_e$ has an even number of thick segments piercing the minimal
surface.  Yet it still has an exponential fall off with area.  
See Fig. 6 which gives the logarithmic derivative showing that $W_e$
has about half the string tension of the $W$.  The thin segments
are still present and they account for the disordering.

\begin{figure}
\epsfig{file=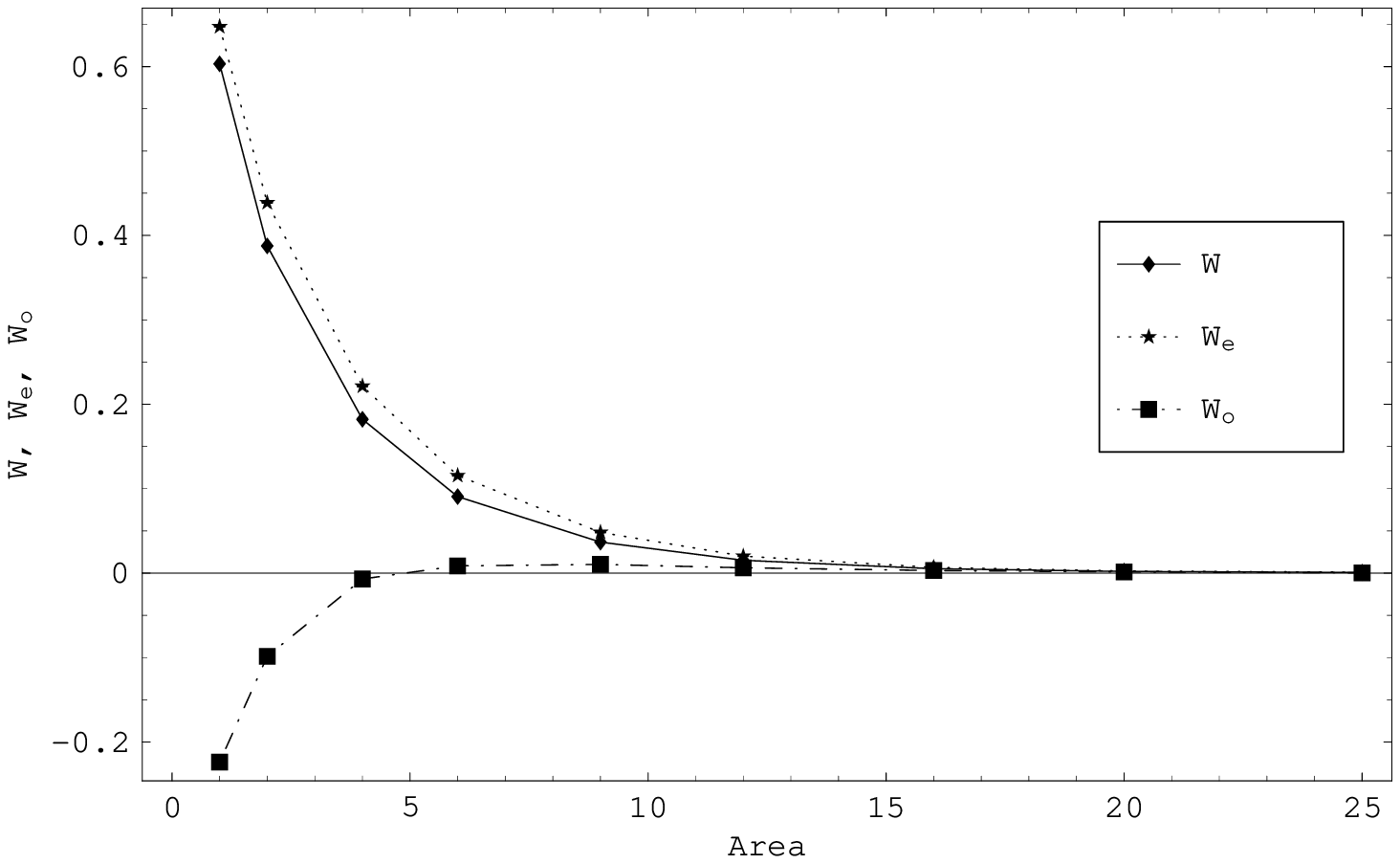,width=15cm}
\epsfig{file=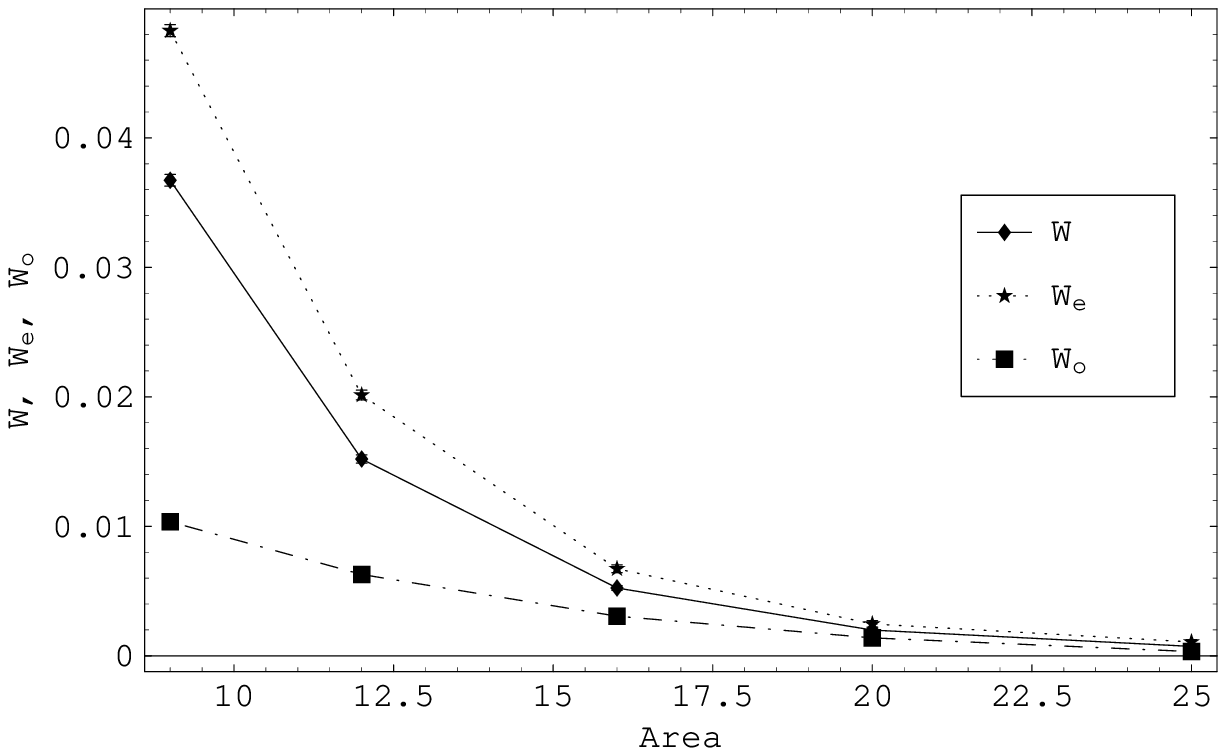,width=15cm}
\center{{\bf Figure 4}: Same as in Figure 2 but tagged by ``thin" only.}
\end{figure}

\item Fig 4:  
[Thin segment piercing the minimal surface] 
By definition $W_e$ has an even number of thin segments piercing the minimal
surface.  It nearly coincides with $W$ and hence cancellations
due to thin segments not important and  are lost in the noise.  
This curve also shows a breakdown of the correlation of the sign of $W_o$ and the sign of the vortex counter for all but the small loop sizes.  This object is not a good predictor of the
sign of the Wilson loop.

\begin{figure}
\epsfig{file=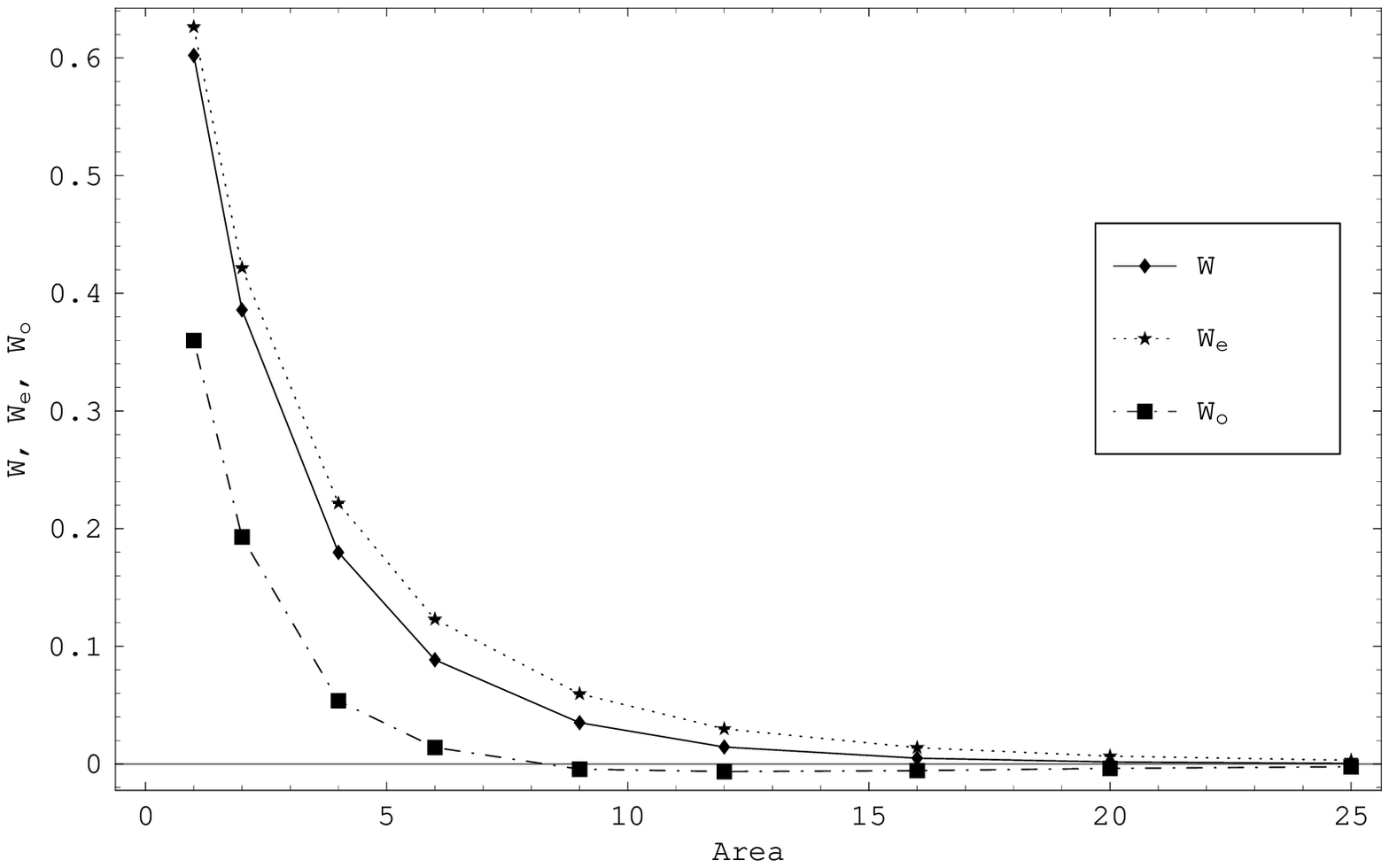,width=15cm}
\epsfig{file=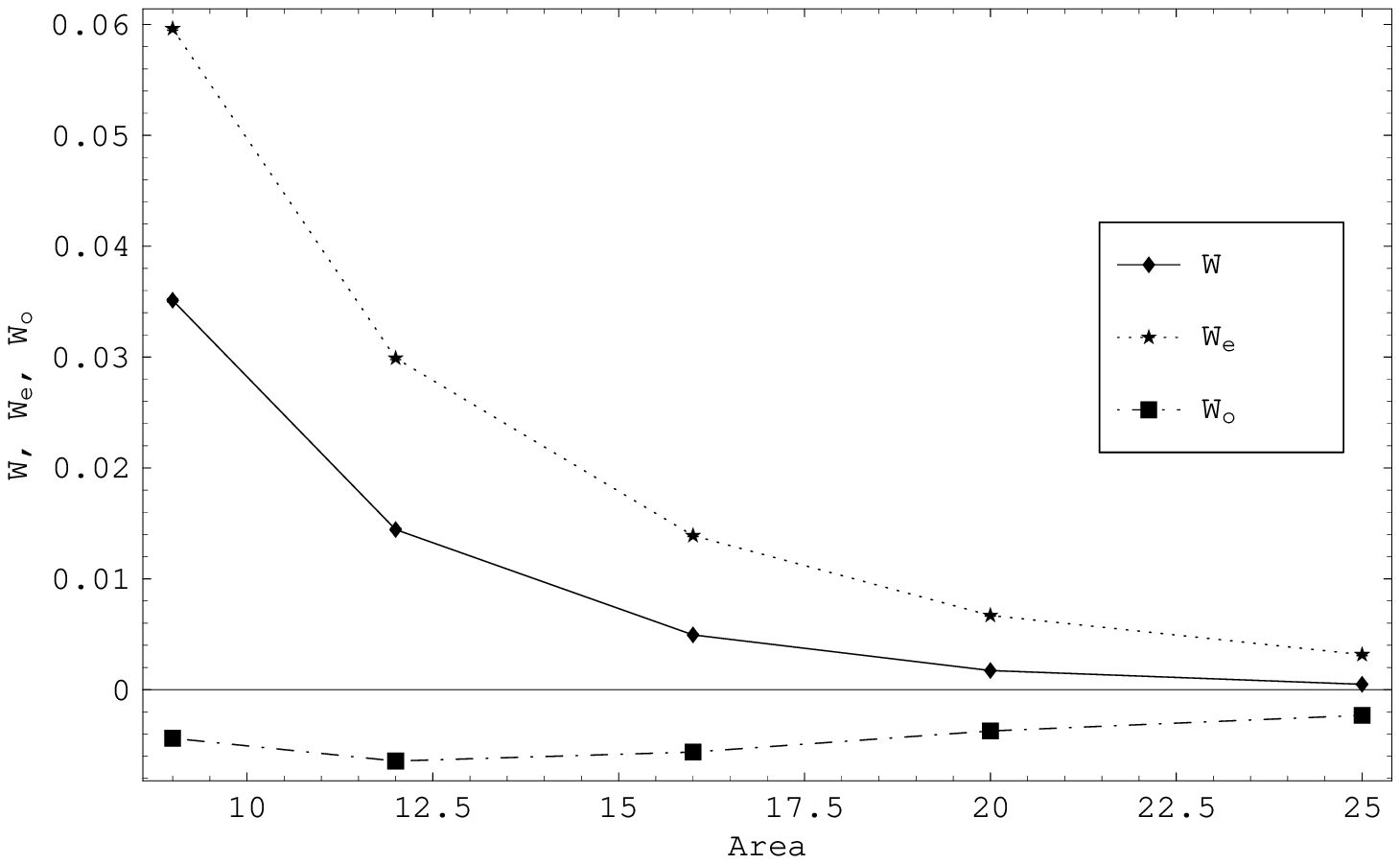,width=15cm}
\center{{\bf Figure 5}: Same as in Figure 2 but tagged by ``projection" only.}
\end{figure}

\item Fig. 5:  [Projection vortices piercing the minimal surface]
 For loops of area 9 and higher, the 
sign of the vortex counter correlates with $W_o$. The last four points coincide with those reported in Ref.\cite{dfgo}.  
Comparing Fig. 3 [thick] and Fig.5b. [projection] the latter data are about a factor of 10 smaller.

\begin{figure}
\epsfig{file=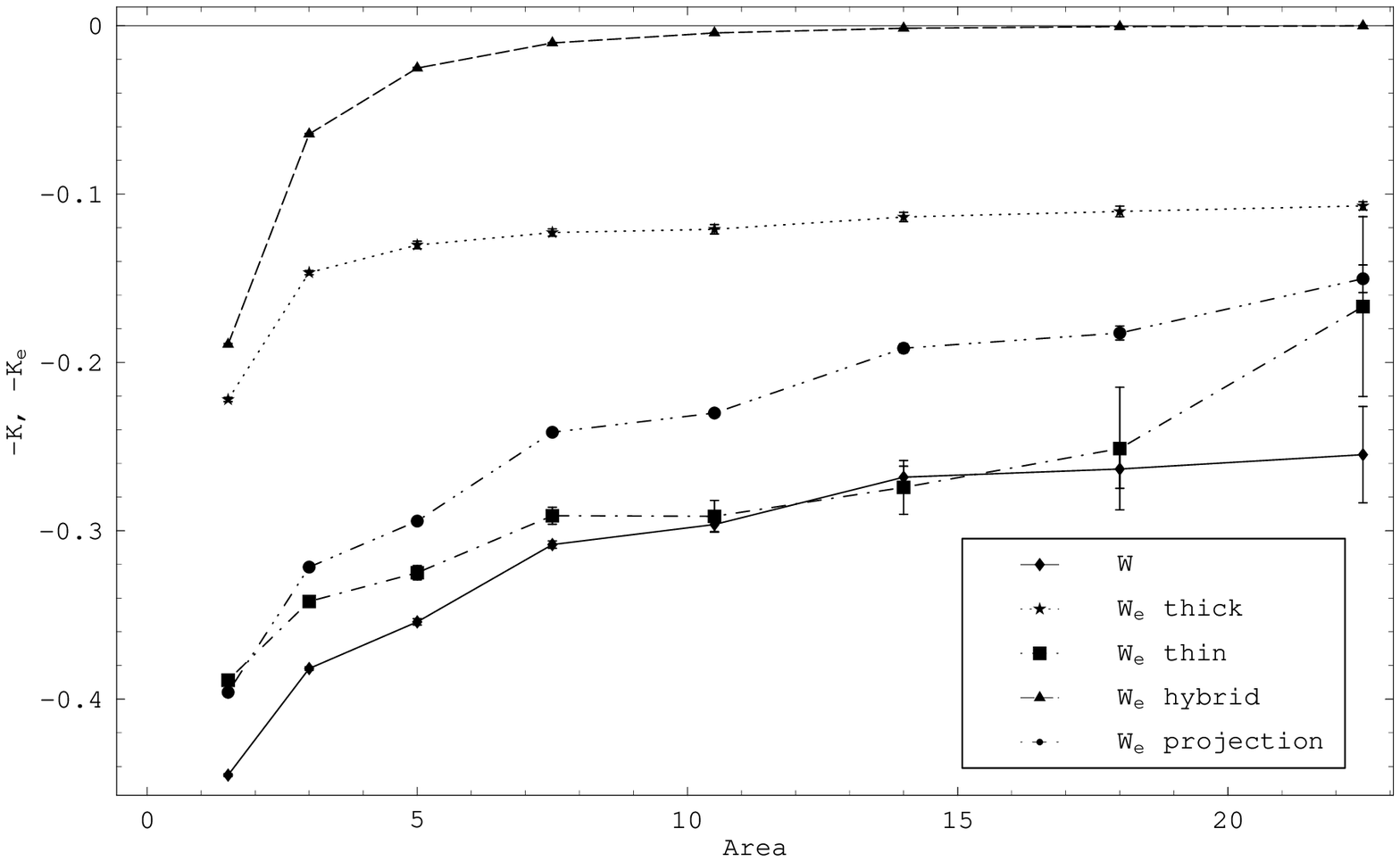,width=15cm}
\center{{\bf Figure 6}: String tension: $K=-\frac{1}{W}\frac{dW}{dArea}$, similarly 
for $K_e$.}
\end{figure}

\item Fig. 6:  [Logarithmic derivatives]
$W$ and $W_e$ [thick] show a constant string tension for larger loops.  We suspect that with better statistics, $W_e$ [thin] will also.  
$W_e$ [hybrid] shows the vanishing of the string tension if one removes
the disordering mechanism completely.  Larger loop areas are needed
to decide if the logarithmic derivative of  $W_e$ [projection] will
also go to zero, or stablize which would indicate that a disordering mechanism remains.

\end{itemize}

\section{Conclusion}

We have proposed a simulation algorithm for the partition function in the 
$Z(2) \times  SO(3)$ formulation.  We show that the algorithm 
is ergodic, reaching all relevant configurations.
We consider vortices which wrap around the torus. 
 We find that these have zero weight in the partition function reflecting
the fact that they are topologically stable. 

Thick vortices are known to be an important factor in disordering the Wilson loop.
As a first calculation, we measure various vortex counters, in order to see how
they are correlated with other observables.

\vspace{0.20in}
\noindent {\bf Acknowledgments}

We are pleased to thank E. T. Tomboulis and S. Cheluvaraja for many 
helpful discussions.   R. H. would like to
thank Tom DeGrand and the Physics Department at the University of Colorado for 
their hospitality where part of this work was done. 
This work was supported in part by United States Department of Energy
grant DE-FG05-91 ER 40617.

\appendix
%
%
\section{\bf Appendix on Z(2) Algebra}
%
%

We follow the definitions of Tomboulis and Kovacs \cite{t,kt}
We used b, p and c to denote the links, plaquettes and cubes respectively. Occasionally we use indices (e.g. $p_{ijkl}^{\mu\nu}$) to denote the objects location ($ijkl$) and orientation ($\mu\nu$). The $\partial$ and $\hat\partial$ operators have the usual meaning: the boundary and coboundary operators.

$Z(2)$ is the multiplicative group with two elements and we will denote it's elements with Greek letters: $\alpha$, $\beta$ ... The $Z(2)$ group admits two representations: $M_+(\pm 1)=+1$ and $M_-(\pm 1)=\pm 1$. The characters of the representations are: $\chi_{+1}(\pm 1)=+1$ for $M_+$ and $\chi_{-1}(\pm 1)=\pm 1$ for $M_-$.

The $Z(2)$ delta function is defined as: $\delta(+1)=1$ and $\delta(-1)=0$. 

We list the basic of the properties of the characters:
\ben
\chi_\sigma (\tau)&=&\chi_\tau (\sigma), \\
\chi_\tau (\alpha\beta)&=&\chi_\tau (\alpha) \chi_\tau(\beta), \\
\delta(\alpha)&=&  \half \sum_{\tau} \chi_\tau (\alpha).
\een

\section{\bf Appendix,  Proof of ergodicity of the $\sigma $ update algorithm}
\label{appendix}

In this appendix we prove that the two groups described in Sec. 4:
  ${\cal D}$ built up by ``star transformations"
on each link is identical to the group 
$\overline{\cal C}$ defined by constraints on each link.  Sections B.1, B.2 and B.3
give preliminaries. Section B.4 proves the result by induction.  We define intermediate
groups ${\cal D}_E$ built out of star transformations on a subset of 
links and similarly  $\overline{\cal C}_E$ restricted by constraints on the same subset
of links.  We then increase the set to $E'$ by an additional link and proceed by 
induction.

The partition function contains the factor, 
Eqns.(\ref{z36}, \ref{tau}, \ref{C}, \ref{CC}, \ref{CCC}, \ref{CCCC})
\ben
C[\alpha]=\sum_{\tau\in{\cal C}}\prod_p \chi_{\tau(p)}(\alpha(p))=\sum_{\tau\in{\cal C}}\la\alpha,\tau\ra.
\een
We will see that $C[\alpha]=0$ for $\alpha\not\in\bar{\cal C}$ and $C[\alpha]=|{\cal C}|$ 
$=$ number of elements in the set ${\cal C}$ for $\alpha\in\bar{\cal C}$, where  
\ben
\bar{\cal C}=\{\alpha\in{\cal C} |\la\alpha,
\tau\ra=1,\;\;\;\;\forall\tau\in{\cal C}\},
\een
and where $\cal C$ is defined to be:
\ben
{\cal C}=\{\tau\in{\cal A}|\prod_b  \delta(\tau(\hat{\partial}b))=1\}
\;\;\;\; \Longleftrightarrow  \;\;\;\;
{\cal C}=\{\tau\in{\cal A}|\tau(\hat{\partial}b)=1,\;\;\;\;\forall b\in B\}.
\een
and where $B$ is the set of all links of the lattice.

 Recall ${\cal C}$ is a closed tiled surface of negative $\tau$
plaquettes as required by the constraints in the partition function. The $\tau$
variables are summed, leaving the $\alpha$ variables.
$\bar{\cal C}$ contains, e.g., a vortex of stacked negative $\alpha$ plaquettes 
which have non-vanishing weight in the partition function.

\subsection{Notation}

$P$ denotes the set of all plaquettes of the lattice and $B$ the set of all links. A ``configuration" is defined to be a function that associates an element of $Z_2$ to each plaquette. We will denote the configurations with Greek letters and write:
\ben
\alpha:P\pe Z_2 \;\;\;\;\; p\mapsto\alpha(p)\in Z_2.
\een
The set of all configurations is denoted by ${\cal A}=\{\alpha:P\pe Z_2\}$. For every set $K$ we will denote with $|K|$ the number of elements in the set. $|{\cal A}|=2^{6N}$ where $N$ is the number of sites in the lattice.

Configurations form a group under multiplication:
let $\alpha$ and $\beta$ be two configurations in $\cal A$. Then $\alpha\beta$ is defined to be:
\ben
\alpha\beta(p)=\alpha(p)\beta(p).
\een
We denote the unit element $1\in{\cal A}$ which assigns $+1$  to all
plaquettes.  Elements are their own inverses.

The partition function, Eqn. \ref{z36}, contains a summation 
 over all $\tau$ configurations that satisfy the constraints: $\prod_{b} \delta \left( \tau(\hat{\partial} b) \right)=1$.  We denote this set by ${\cal C}$:
\be
C[\alpha]=\sum_{\tau\in {\cal C}} \prod_p \chi_{\tau(p)}(\alpha(p)),
\ee
where ${\cal C}=\{\tau\in{\cal A}|\prod_{b} \delta \left( \tau(\hat{\partial} b) \right)=1\}$. 

Simplifying the notation in the summand we define:
\ben
\la\tau,\alpha\ra  \equiv \prod_{p}\chi_{\tau(p)}(\alpha(p)).
\een
We will list here some properties of the bracket $\la\cdot,\cdot\ra$ without proof. Lets take $\alpha,\beta,\gamma\in{\cal A}$. Then:
\ben
\la\alpha,\beta\ra&=&\la\beta,\alpha\ra, \\
\la\alpha\beta,\gamma\ra&=&\la\alpha,\gamma\ra\la\beta,\gamma\ra, \\
\la\alpha,\beta\gamma\ra&=&\la\alpha,\beta\ra\la\alpha,\gamma\ra, \\
\la\alpha,1\ra&=&\la1,\alpha\ra=1.
\een

\subsection{Some theorems regarding the subgroups of ${\cal A}$}

Consider an arbitrary subgroup  of ${\cal A}$ denoted ${\cal K}$,
(of which ${\cal C}$ is an example).

{\bf Proposition 1}: Let $\cal K$ be a subgroup of $\cal A$ and define:
\ben
K[\alpha]=\sum_{\beta\in{\cal K}}\la\alpha,\beta\ra=\sum_{\beta\in{\cal K}}\la\beta,\alpha\ra.
\een
Then:
\be
K[\alpha]=\la\alpha,\beta_0\ra K[\alpha]
\ee
for any $\beta_0\in{\cal K}$.

{\bf Proof}: Since ${\cal K}$ is a group we have:
\ben
\sum_{\beta\in{\cal K}}\la\alpha,\beta_0\beta\ra=\sum_{\beta\in{\cal K}}\la\alpha,\beta\ra
\een
where we used the property of the group sum 
 that $\{ \beta_0\beta \} $ is a rearrangement of the group elements
$\{ \beta \}$.   Using the properties of the bracket we have:
\ben
K[\alpha]&=&\sum_{\beta\in{\cal K}}\la\alpha,\beta_0\beta\ra=\sum_{\beta\in{\cal K}}\la\alpha,\beta_0\ra\la\alpha,\beta\ra, \\
&=&\la\alpha,\beta_0\ra\sum_{\beta\in{\cal K}}\la\alpha,\beta\ra=\la\alpha,\beta_0\ra K[\alpha].
\een

{\bf Definition}: Let $\cal K$ be a subgroup of $\cal A$. We define:
\ben
\bar{\cal K}=\{\alpha\in{\cal A}|\la\alpha,\beta\ra=1 \;\;\;\;\forall\beta\in{\cal K}\}.
\een
$\bar{\cal K}$  (e.g. $\overline{\cal C}$) has always at least one element, $1$, since $\la\alpha,1\ra=1$ for any $\alpha\in\cal A$.  Moreover it is easy to prove that $\bar{\cal K}$ is a group too. Now we note the following lemma:

{\bf Lemma 1}: $\bar{\cal A}=\{1\}$.

{\bf Proof}: Choose $\alpha\in\cal A$ with $\alpha\not=1$. This means that there is at least one plaquette $p_0\in P$ for which $\alpha(p_0)=-1$. Then if we take $\beta(p)=1$ for all $p\not=p_0$ and $\beta(p_0)=-1$ which is an element of $\cal A$ we see that $\alpha$ and $\beta$ have only one plaquette, $p_0$, on which both are $-1$. Then
\ben
\la\alpha,\beta\ra=\prod_p \chi_{\alpha(p)}(\beta(p))=-1,
\een
proving that $\alpha\not\in\bar{\cal A}$. Thus we proved that if $\alpha\not=1$ then $\alpha\not\in\bar{\cal A}$. Hence $\bar{\cal A}=\{1\}$.

{\bf Proposition 2}: Let $\cal K$ be a subgroup of $\cal A$. If $K[\alpha]$ is defined as in proposition 1 we have $K[\alpha]=0$ for $\alpha\not\in\bar{\cal K}$ and $K[\alpha]=|K|$ for $\alpha\in\bar{\cal K}$.

{\bf Proof}: Using proposition 1 we have:
\ben
K[\alpha]=\la\alpha,\beta_0\ra K[\alpha]
\een
for any $\beta_0\in\cal K$. Clearly if $\la\alpha,\beta_0\ra\not=1$ then $K[\alpha]=0$. Thus $K[\alpha]=0$ for all $\alpha$ that have at least one element $\beta_0\in\cal K$ for which $\la\alpha,\beta_0\ra\not=1$. This is equivalent with saying that if $\alpha\not\in\bar{\cal K}$ then $K[\alpha]=0$, thus proving the first part.

Now consider $\alpha\in\bar{\cal K}$. This means that $\la\alpha,\beta\ra=1$ for all $\beta\in\cal K$. Then:
\ben
K[\alpha]=\sum_{\beta\in{\cal K}}\la\alpha,\beta\ra=\sum_{\beta\in{\cal K}} 1=|{\cal K}|.
\een
This concludes the proof.

{\bf Theorem 1}: Let $\cal K$ be a subgroup of $\cal A$. Then $|{\cal K}||\bar{\cal K}|=|{\cal A}|$.

{\bf Proof}: Consider the following:
\ben
I=\sum_{\alpha\in{\cal A}}K[\alpha]=|{\cal K}|\sum_{\alpha\in\bar{\cal K}} 1=|{\cal K}||\bar{\cal K}|,
\een
where we used proposition 2. Interchange the sums in $I$ and using the commutative
property of the bracket:
\ben
I=\sum_{\alpha\in{\cal A}}K[\alpha]=\sum_{\alpha\in{\cal A}}\sum_{\beta\in{\cal K}}\la\alpha,\beta\ra=\sum_{\beta\in{\cal K}}\sum_{\alpha\in{\cal A}}\la\beta,\alpha\ra=\sum_{\beta\in{\cal K}} A[\beta].
\een
where $A[\beta]$ is defined exactly as $K[\alpha]$ using the fact that the bracket is commutative. Now proposition 2 tells us that $A[\alpha]=|{\cal A}|$ for $\alpha\in\bar{\cal A}$ and 0 otherwise. Using lemma 1 we have that $A[1]=|{\cal A}|$ and zero otherwise and thus we get $I=|{\cal A}|$. But since we already know that $I=|{\cal K}||\bar{\cal K}|$ we have:
\ben
|{\cal K}||\bar{\cal K}|=|{\cal A}|.
\een

\subsection{Constructing ${\cal D}$  }

We already know that $\bar{\cal C}$ is a subgroup of $\cal A$. In this section we construct a subgroup of $\bar{\cal C}$, denoted ${\cal D}$, which, in the next section, will be shown to be the entire $\bar{\cal C}$.

{\bf Definition}: We will call ``a star configuration around link b" the following configuration:
\ben
\stb(p)=\left\{ \matrix{+1&p\not\in\hat{\partial}b, \cr -1&p\in\hat{\partial}b.} \right.
\een

{\bf Lemma 2}: For any link b we have $\stb \in \overline{{\cal C}}$.

{\bf Proof}: We can prove that $\stb$ is a member of
 $\bar{\cal C}$ by checking that:
\ben
\la\stb,\tau\ra=1,\;\;\;\;\forall \tau\in{\cal C}.
\een
Consider any link b and any $\tau\in{\cal C}$ and compute:
\ben
\la\stb,\tau\ra=\prod_p \chi_{\stb(p)}(\tau(p))=\prod_{p\not\in\hat{\partial}b} \chi_1(\tau(p))\prod_{p\in\hat{\partial}b} \chi_{-1}(\tau(p))=\chi_{-1}(\tau(\hat{\partial}b)).
\een
Since $\tau\in{\cal C}$ then $\tau(\hat{\partial}b)=1$ for any link b. 
Then:
\ben
\la\stb,\tau\ra=\chi_{-1}(\tau(\hat{\partial}b))=\chi_{-1}(1)=+1.
\een
for all $\tau\in{\cal C}$. This proves that for any link b we have $\stb\in\bar{\cal C}$.

If we take the set of all  star configurations by taking all possible products between them we can generate a group, $\cal D$. Since all the star configurations are included in $\bar{\cal C}$, which is a group itself, the group $\cal D$ is a subgroup of $\bar{\cal C}$. Lets write $\cal D$ explicitely:
\ben
{\cal D}=\{\alpha\in{\cal A}|\alpha=\prod \stbi \}.
\een 
where the product is over all configurations reached by star transformations.
  In order to prove that $\cal D$ covers all $\bar{\cal C}$ we will show that $\cal D$ has the same number of elements as $\bar{\cal C}$. Since ${\cal D}\subseteq\bar{\cal C}$ we have $|{\cal D}|\leq|\bar{\cal C}|$. All we have to prove now is that $|{\cal D}|\geq|\bar{\cal C}|$.

\subsection{Proof that  $\bar{\cal C} \equiv  {\cal D}$ }

We now prove that $|{\cal D}|\geq|\bar{\cal C}|$. Using theorem 1 we know that $|{\cal C}||\bar{\cal C}|=|{\cal A}|$.  
Thus we can prove that $|{\cal D}|\geq|\bar{\cal C}|$ by proving that $|{\cal C}||{\cal D}|\geq|{\cal A}|$. Lets recall
the definitions of these sets:
\ben
{\cal C}&=&\{\tau\in{\cal A}|\tau(\hat{\partial}b)=1,\;\;\;\;\forall b\in B\}, \\
{\cal D}&=&\{\alpha\in{\cal A}|\alpha=\prod \stbi,\;\;\;\; b_i\in B\}.
\een

As we see the set $\cal C$ is constructed by means of eliminating the configurations that do not obey a certain constraint whereas the set $\cal D$ is constructed by generating all possible combinations built from star configurations acting on the identity configuration. 
These sets admit a generalization as follows. Define:
\ben
{\cal C}_E&=&\{\tau\in{\cal A}|\tau(\hat{\partial}b)=1,\;\;\;\;\forall b\in E\}, \\
{\cal D}_E&=&\{\alpha\in{\cal A}|\alpha=\prod \stbi,\;\;\;\; b_i\in E \},
\een
where $E\subseteq B$. It is easy to check that both ${\cal C}_E$ and ${\cal D}_E$ are groups. In words, ${\cal C}_E$ is the set of all the configurations that obey the constraint only on the subset $E$ of all links and ${\cal D}_E$ is the group generated by star configurations associated only with links in $E$. It is obvious that ${\cal C}_B={\cal C}$ and that ${\cal D}_B={\cal D}$. If we prove that $|{\cal D}_E||{\cal C}_E|\geq|{\cal A}|$ we will implicitly prove that $|{\cal D}||{\cal C}|\geq|{\cal A}|$ and thus proving that $|{\cal D}|\geq|\bar{\cal C}|$. 

{\bf Theorem 2}: $|{\cal D}_E||{\cal C}_E|\geq|{\cal A}|$.

{\bf Proof}: We will prove this using induction. The first step will be to prove this for $E=\emptyset$ and the second step will be to prove that the relation holds for $E\purr=E\cup\{b\}$, where b is any link, assuming that the inequality holds for $E$.

\subsubsection{Initializing}
 Let $E=\emptyset$. We have ${\cal C}_\emptyset={\cal A}$ since there is no constraint that has to be obeyed and ${\cal D}_\emptyset=\{1\}$ since this is the group that has no star configuration in it. Then we have:
\ben
|{\cal D}_\emptyset||{\cal C}_\emptyset|=1\times|{\cal A}|=|{\cal A}|.
\een
Therefore the inequality is verified for $E=\emptyset$.

\subsubsection{Iterative inductive step}
 Assume that:
\ben
|{\cal D}_E||{\cal C}_E|\geq|{\cal A}|
\een
for a certain $E$. Let us prove that this also holds for $E\purr=E\cup\{b\}$ for any $b\not\in E$. We will see what happens with ${\cal C}_E$ and ${\cal D}_E$ when we increase $E$ by one element.

{\bf Iteration on ${\cal C}_E$}

When we increase the number of elements in $E$, ${\cal C}_E$ grows smaller since there will be more constraints to obey. It may happen that the new constraint, namely $\tau(\hat{\partial}b)=1$, is superfluous i.e.  all configuration in ${\cal C}_E$ already satisfy this constraint. In this case ${\cal C}_{E\purr}={\cal C}_E$ and $|{\cal C}_{E\purr}|=|{\cal C}_E|$. Lets now see what happens if there is at least one element in ${\cal C}_E$ that doesn't obey the new constraint. In this case we can break down ${\cal C}_E$ in two disjoint sets: ${\cal C}_{E\purr}$ the sets of all configuration $\tau\in{\cal C}_E$ that obey the new constraint and $R$ the set of all configurations $\tau\in{\cal C}_E$ that do not obey the constraint. In case that $R\not=\emptyset$ lets take $\tau_0\in R$. Using this we can construct a one to one mapping between ${\cal C}_{E'}$ and $R$:
\ben
f:{\cal C}_{E\purr}\pe R,\;\;\;\; f(\tau)=\tau_0\tau, \\
f^{-1}:R\pe {\cal C}_{E\purr},\;\;\;\; f(\tau)=\tau_0\tau.
\een
It is easy to check that $f$ is indeed one to one.

Now since there is a one to one mapping between ${\cal C}_{E\purr}$ and $R$ we have that $|{\cal C}_{E\purr}|=|R|$. But since they are disjoint sets of ${\cal C}_E$ and ${\cal C}_{E\purr}\cup R={\cal C}_E$ we have $|{\cal C}_{E\purr}|+|R|=|{\cal C}_E|$. Thus we have $|{\cal C}_{E\purr}|=\half |{\cal C}_E|$.

Summing up we know that by adding a new constraint $|{\cal C}_{E\purr}|$ is either equal with $|{\cal C}_E|$ when the constraint is superfluous or is $\half |{\cal C}_E|$ when we have at least one element in ${\cal C}_E$ that violates the constraint.

{\bf Iteration on ${\cal D}_E$}

Now we will look at ${\cal D}_{E\purr}$. What happens when we pass from ${\cal D}_E$ to ${\cal D}_{E\purr}$? We add a new element in the group. Thus the group grows larger with one exception: it may happen that $\stb\in {\cal D}_E$ although $b\not\in E$. Then the group will stay the same since every combination that doesn't involve $\stb$ is already in the group ${\cal D}_E$ and every combination that involves $\stb$ can be generated by elements already in the group. Formally, if $\stb\in{\cal D}_E$ then for any $\alpha\in {\cal D}_E$ we have $\stb\alpha\in{\cal D}_E$ and thus ${\cal D}_{E\purr}={\cal D}_E$. This proves that if $\stb\in{\cal D}_E$ we have $|{\cal D}_{E\purr}|=|{\cal D}_E|$. It is not obvious that this case happens but we can easily construct one: take a site in the lattice and take the eight links that form its coboundary. Since the coboundary of a coboundary is nul (i.e. ${\hat\partial}^2=0$) it means that if we perform all the star transformations associated with the links in the coboundary of the site we get the identity. This means that the product of any seven of this star transformations is equal to the eighth star transformation.

Consider what happens when $\stb\not\in {\cal D}_E$. Then for every element $\alpha\in{\cal D}_E$ we have two elements in ${\cal D}_{E\purr}$: $\alpha\in{\cal D}_{E\purr}$ and $\stb\alpha\in{\cal D}_{E\purr}$. We can actually construct two sets in ${\cal D}_{E\purr}$: ${\cal D}_E$ which is trivially included in ${\cal D}_{E\purr}$ and $\stb\times {\cal D}_E=\{\alpha\in{\cal D}_{E\purr}|\alpha=\stb\beta,\;\;\;\beta\in{\cal D}_E\}$. If $\stb\not\in{\cal D}_E$ then these two sets do not overlap.

If they were to overlap then there is $\alpha\in{\cal D}_E\cap \stb\times{\cal D}_E$ with $\alpha\in{\cal D}_E$ and $\alpha=\stb\beta$ with $\beta\in{\cal D}_E$. Then $\stb=\alpha\beta$ and since both $\alpha$ and $\beta$ are members of ${\cal D}_E$ then $\stb\in{\cal D}_E$ which contradicts our assumption.

It is easy to see that ${\cal D}_E$ and $\stb\times{\cal D}_E$ have the same number of elements and that ${\cal D}_E\cup \stb\times{\cal D}_E={\cal D}_{E\purr}$. Now since ${\cal D}_E\cup \stb\times{\cal D}_E={\cal D}_{E\purr}$ and ${\cal D}_E\cap \stb\times{\cal D}_E=\emptyset$ we have $|{\cal D}_E|+ |\stb\times{\cal D}_E|=|{\cal D}_{E\purr}|$. Moreover $|{\cal D}_E|=|\stb\times{\cal D}_E|$ and thus $|{\cal D}_{E\purr}|=2|{\cal D}_E|$.

Summing up if $\stb\in{\cal D}_E$ then $|{\cal D}_{E\purr}|=|{\cal D}_E|$ and if $\stb\not\in{\cal D}_E$ we have $|{\cal D}_{E\purr}|=2|{\cal D}_E|$.

{\bf Intermediate summary of possible cases}

Thus we arrived at the conclusion that if we add another link $b$ to $E$ one of the four following things may happen:
\begin{itemize}
\item{the new constraint is superfluous ($|{\cal C}_{E\purr}|=|{\cal C}_E|$)  and $\stb\not\in{\cal D}_E$ ($|{\cal D}_{E\purr}|=2|{\cal D}_E|$). Then $|{\cal C}_{E\purr}||{\cal D}_{E\purr}|=2|{\cal C}_E||{\cal D}_E|\geq|{\cal A}|$.}
\item{the new constraint is superfluous ($|{\cal C}_{E\purr}|=|{\cal C}_E|$)  and $\stb\in{\cal D}_E$ ($|{\cal D}_{E\purr}|=|{\cal D}_E|$). Then $|{\cal C}_{E\purr}||{\cal D}_{E\purr}|=|{\cal C}_E||{\cal D}_E|\geq|{\cal A}|$.}
\item{the new constraint is not superfluous ($|{\cal C}_{E\purr}|=\half|{\cal C}_E|$)  and $\stb\not\in{\cal D}_E$ ($|{\cal D}_{E\purr}|=2|{\cal D}_E|$). Then $|{\cal C}_{E\purr}||{\cal D}_{E\purr}|=|{\cal C}_E||{\cal D}_E|\geq|{\cal A}|$.}
\item{the new constraint is not superfluous ($|{\cal C}_{E\purr}|=\half |{\cal C}_E|$)  and $\stb\in{\cal D}_E$ ($|{\cal D}_{E\purr}|=|{\cal D}_E|$). Then $|{\cal C}_{E\purr}||{\cal D}_{E\purr}|=\half|{\cal C}_E||{\cal D}_E|$ and we don't know if this is smaller or greater than $|{\cal A}|$.}
\end{itemize}
If we look at the scheme above we see that if we can prove that the last case never happens than we proved our theorem. This is exactly what we prove in the following lemma:

{\bf Lemma 3}: If $\stb\in{\cal D}_E$ then the new constraint is superfluous.

{\bf Proof}: If $\stb\in{\cal D}_E$ then there is a subset $E_0\subseteq E$ with the property $\stb=\prod_{b'\in E_0} \stbp$ (this simply asserts that $\stb$ can be written as a product of star configurations associated with a subset of links  in $E$). 

Lets denote with $P_0$ the set of all plaquettes that form the coboundary of $E_0$: $P_0=\hat{\partial}E_0$ 
All the plaquettes in $P_0$ have at least one neighboring link in $E_0$ and thus they are flipped at least once when you take the product $\prod_{b'\in E_0} \stbp$. 

Since the whole product is equal to $\stb$ the coboundary of $b$, $\hat{\partial}b$, has to be included in $P_0$. Otherwise we cannot flip the plaquettes around $b$ and we cannot form $\stb$. 

Further since the final state is $\stb$ we 
know that all plaquettes $p\in P_0-\hat{\partial}b$ are flipped an even number of times and all plaquettes $p\in\hat{\partial}b$ are flipped an odd number of times. 

We define a function $\kappa(p)$ on plaquettes in $P_0$ that returns the number of times this plaquette is flipped. 
Using this it is easy to prove that $\kappa(p)$ is even for $p\in P_0-\hat{\partial} b$ and odd for $p\in\hat{\partial} b$:
\ben
\prod_{b'\in E_0}\stbp(p)=
\prod_{b'\in E_0}(-1)^{\epsilon_{b'}(p)}=(-1)^{\sum_{b'\in E_0}
\epsilon_{b'}(p)}=(-1)^{\kappa(p)}
\een
where $\epsilon_{b'}(p)$ is 1 if $p\in\hat{\partial}b'$ and zero otherwise. Now since $\left[\prod_{b'\in E_0}\stbp\right]=\stb$ and since $\stb=(-1)^{\epsilon_{b}(p)}$ we have that $\kappa(p)$ and $\epsilon_p(b)$ have to have the same parity proving that $\kappa(p)$ is even for $p\in P_0-\hat{\partial} b$ and odd for $p\in\hat{\partial} b$.

Now lets take an element $\tau\in{\cal C}_E$. Since $\tau(\hat{\partial}b')=1$ for all $b'\in E$ and $E_0\subseteq E$ we have that:
\ben
\prod_{b'\in E_0}\tau(\hat{\partial}b')=1
\een
At the same time we can write:
\ben
\prod_{b'\in E_0}\tau(\hat{\partial}b')=\prod_{p\in P_0}(\tau(p))^{\kappa(p)}=1
\een
But since $\kappa(p)$ is even for $p\in P_0-\hat{\partial} b$ and odd for $p\in\hat{\partial}b$ we have:
\ben
\prod_{p\in P_0}(\tau(p))^{\kappa(p)}=\prod_{p\in \hat{\partial} b}\tau(p)=\tau(\hat{\partial}b)=1
\een
Thus we proved that if $\tau\in{\cal C}_E$ then $\tau(\hat{\partial}b)=1$ and thus the new constraint is superfluous. This proves the lemma.

Now using this lemma we see that the case where $|{\cal C}_{E\purr}|=\half |{\cal C}_E|$ and $|{\cal D}_{E\purr}|=|{\cal D}_E|$ never happens and since in all the other cases our theorem holds then we proved our theorem.

\subsection{Summary}

We proved using theorem 2 that $|{\cal D}_E||{\cal C}_E|\geq|{\cal A}|$ 
which means that $|{\cal D}||{\cal C}|\geq|{\cal A}|$. Using this and 
the result of theorem 1: $|\bar{\cal C}||{\cal C}|=|{\cal A}|$ we see that 
$|{\cal D}|\geq|\bar{\cal C}|$ but we already know that 
since ${\cal D}\subseteq\bar{\cal C}$ we have that $|{\cal D}|\leq|\bar{\cal C}|$. The only possible solution to this is that $|{\cal D}|=|\bar{\cal C}|$ and since ${\cal D}\subseteq\bar{\cal C}$ we have that ${\cal D}=\bar{\cal C}$.

This is the result that we were looking for. Now we can say that:
\be
C[\alpha]=\left\{\matrix{|{\cal C}|\;\;\;\;\alpha\in{\cal D} \cr 0\;\;\;\;\alpha\not\in{\cal D}}\right.
\ee
where $\cal D$ is the group formed by taking all possible products between the star configurations.

\section{\bf Appendix,  Antiperiodic boundary conditions}
\label{appendixc}

We have shown that vortices wrapped around the periodic boundary conditions have
zero weight.  However we point out how these configurations can instead be
weight $= 1$ and the above configurations weight $= 0$ by a minor 
change in the formulation. Consider 
\be
Z=\int [dU] 
e^{
\left(
   \beta 
   \sum_{p} 
   \half \lab{tr} [  U(\partial p)]  
\right)
}
\rightarrow 
\int [dU] 
e^{
\left(
   \beta 
   \sum_{p\not\in P} 
   \half \lab{tr} [  U(\partial p)]  
   -
   \beta 
   \sum_{p\in P} 
   \half \lab{tr} [  U(\partial p)]  
\right)
}.
\label{abc}
\ee
where $P$ is a co-plane of sign flipped terms in the action.  This is known as
antiperiodic  or twisted boundary conditions in the literature. 

Antiperiodic boundary conditions amounts to nothing more than a change in the action.
The derivation of the partition function in terms of $SU(2)/Z_2$ and $Z_2$ variables can be generalized. In our notation , Eqn.(\ref{abc}) becomes
\be
Z &=& \int [dU] 
\sum_{\sigma \eta \in{\bar{\cal C}}}
e^{
\left(
   \beta 
   \sum_{p} 
   \half |\lab{tr} [  U(\partial p)]  | \sigma(p)
\right)
}
\rightarrow 
\nonumber
\\
&&\int [dU] 
\sum_{\sigma\eta\in{\bar{\cal C}}}
e^{
\left(
   \beta 
   \sum_{p\not\in P} 
   \half |\lab{tr} [  U(\partial p)]  | \sigma(p)
   -
   \beta 
   \sum_{p\in P} 
   \half |\lab{tr} [  U(\partial p)]  | \sigma(p)
\right)
}.
\label{abcp}
\ee

Define:
\ben
\sigma'=\sigma_0 \sigma\Leftrightarrow \sigma=  \sigma_0 \sigma'
\een
where $\sigma_0(p)=-1$ for  $p\in P$ and $+1$ elsewhere.

We can simplify $Z$,
Eqn.(\ref{abcp})
\ben
Z &=& \int [dU] 
\sum_{\sigma' \sigma_0 \eta \in{\bar{\cal C}}}
e^{
\left(
   \beta 
   \sum_{p} 
   \half |\lab{tr} [  U(\partial p)]  | \sigma'(p)
\right)
}
 = 
 \int [dU] 
\sum_{\sigma'  \eta \in \sigma_0{\bar{\cal C}}}
e^{
\left(
   \beta 
   \sum_{p} 
   \half |\lab{tr} [  U(\partial p)]  | \sigma'(p)
\right)
}
.
\label{abcpp}
\een
This looks exactly like the partition function for the periodic boundary conditions except that  $\eta\sigma'\in \sigma_0 {\bar{\cal C}}$ instead of ${\bar{\cal C}}$.


\begin{thebibliography} {99}
%
\bibitem{y} L. G. Yaffe, Phys. Rev. {\bf D 21}, 1574 (1980).
%
\bibitem{t} E. Tomboulis, Phys. Rev. {\bf D 32}, 2371 (1981).
%
\bibitem{kt} T. G. Kovacs and E. Tomboulis  Phys. Rev. {\bf D 57}, 4054 (1998).
%
\bibitem{mp} G. Mack and V. B. Petkova, Annals of Physics {\bf 123}, 442 (1979);
{\bf 125}, 117 (1980); Z. Phys. {\bf C 12}, 177 (1982).
%
\bibitem{hs} G. Halliday and A. Schwimmer, Phys. Lett. {\bf B 101}, 327 (1981); 
{\bf B 102}, 337 (1981).
%
\bibitem{yo} T. Yonewa   Nucl. Phys. {\bf B 203} [FS5], 130 (1982).
%
\bibitem{c}  J. M. Cornwall, Phys. Rev. {\bf D 26}, 1453 (1979)
%
\bibitem{ttt} T. G. Kovacs and E. Tomboulis hep-lat/0002004, 9912051, 9908031,
Phys. Lett. {\bf B 463} 104 (1999); Nucl. Phys. B, Proc. Suppl {\bf 73}566 (1999);
Phys. Lett. {\bf 443} 239, (1998); J. Math. Phys. {\bf 40}, 4677 (1999).
%
\bibitem{dfgo} L. Del Debbio, M. Faber, J. Greensite and  S. Olejnik, Phys. Rev. 
{\bf D 55}, 2298 (1997), hep-lat/9802003,
%
\bibitem{fgo} M. Faber, J. Greensite and  S. Olejnik, JHEP 9901:008,1999; 
JHEP 9912:012,1999; hep-lat/9911006; hep-lat/9912002
%
\bibitem{agg} J. Ambjorn, J. Giedt, J. Greensite, hep-lat/9907021, hep-lat/9908020
%
\bibitem{lr} K. Langfeld and H. Reinhardt, Phys. Rev. {\bf D 55},7993 (1997)
%
\bibitem{lrt} K. Langfeld, H. Reinhardt and O. Tennert, Phys. Lett. {\bf B 419},
317 (1998);
%
\bibitem{elrt} M. Engelhardt, K. Langfeld, H. Reinhardt and 
O. Tennert, Phys. Lett. {\bf B 431}, 141 (1998); {\bf 452}, 301 (1999); Phys. Rev.
{\bf D 61}:054504,2000; hep-lat/9908026
%
\bibitem{afe} P. de Forcrand and M. D'Elia, hep-lat/9907028; hep-lat/9909005.
%
\bibitem{montero} A. Montero, hep-lat/9907024.
%
\bibitem{stephenson} P.  W. Stephenson, hep-lat/9909022.
%
\bibitem{private} We wish to acknowledge a private communication with E.T. Tomboulis
on this issue.
%
\bibitem{ach} A. Alexandru, S. Cheluvaraja and R. W. Haymaker, in progress.


\end{thebibliography}
\end{document}